\documentstyle[aps,prd, preprint]{revtex}

\def\kappa{\mu }
\begin{document}
\title{Hydrogen atom in the gravitational fields of topological defects}
\author{Geusa de A. Marques\thanks{E-mail:gmarques@fisica.ufpb.br}
 and Valdir B. Bezerra
\thanks{E-mail:valdir@fisica.ufpb.br}.}
\maketitle

\begin{center}
{\it {Departamento de F\'{\i}sica, Universidade Federal da Para\'{\i}ba,}}

{\it {Caixa Postal 5008, 58059-970, Jo\~{a}o Pessoa, PB, Brazil.}}
\end{center}

\abstract

We consider a hydrogen atom in the background spacetimes generated by an
infinitely thin cosmic string and by a point-like global monopole. In both
cases, we find the solutions of the corresponding Dirac equations and we
determine the energy levels of the atom. We investigate how the geometric
and topological features of these spacetimes leads to shifts in the energy
levels as compared with the flat Minkowski spacetime.


\vskip 3.0 cm

\centerline{\bf{I. Introduction} }

The study of quantum systems in curved spacetimes goes back to the end of
twenties and to the beginning of thirties of the last century\cite{a}, when
the generalization of the Schr\"{o}dinger and Dirac equations to curved
spaces has been discussed, motivated by the idea of constructing a theory
which combines quantum physics and general relativity.

Spinor fields and particles interacting with  gravitational fields has
been the subject of many investigations. Along this line of research we can
mention those concerning the determination of the renormalized vacuum
expectation value of the energy-momentum tensor and the problem of creation
of particles in expanding Universes\cite{a1}, and those connected with
quantum mechanics in different background spacetimes\cite{a2} and, in
particular, the ones which consider the hydrogen atom[4-8] in an arbitrary
curved spacetime.

The study of the single-particle states which are exact solutions of the
generalized Dirac equation in curved spacetimes constitutes an important
element to construct a theory that combines quantum physics and gravity and
for this reason, the investigation of the behaviour of relativistic
particles in this context is of considerable interest.

It has been known that the energy levels of an atom placed in a gravitational
field will be shifted as a result of the interaction of the atom with
spacetime curvature[4-8].
These shifts in the energy levels, which would depend on the features of the
spacetime, are different for each energy level, and thus are distinguishable
from the Doppler effect and from the gravitational and cosmological
redshifts, in which cases these shifts would be the same for all spectral
lines. In fact, it was already shown that in the Schwarzschild geometry, the
shift in the energy level due to gravitational effects is different from the
Stark and Zeeman effects, and therefore, it would be possible, in principle,
to separate the shifts in the energy levels caused by electromagnetic and by
gravitational perturbations\cite{3}. Thus, in these  situations the
energy spectrum carries
unambiguous information about the local features of the background spacetime
in which the atomic system is located.

The general theory of relativity, as a metric theory, predict that
gravitation is manifested as curvature of spacetime. This curvature is
characterized by the Riemann tensor. On the other hand, we know that there
are connections between topological properties of the space and local
physical laws in such a way that the local intrinsic geometry of the space
is not sufficient to describe completely the physics of a given system. As an
example of a gravitational effect of topological origin, we can mention
the fact that only when a particle is transported around a cosmic
 string along a closed curve the string is noticed at
all. This situation corresponds to the gravitational analogue\cite{14} of
the electromagnetic Aharonov-Bohm effect\cite{10a}, in which electrons are
beamed past a solenoid containing a magnetic field. These effects are of
topological origin rather than local. In fact, the nontrivial topology of
spacetime, as well as its curvature, leads to a number of interesting
gravitational effects. Thus, it is also important to
investigate the role played by a nontrivial topology, for example, on a
quantum system. As examples of these investigations we can mention the study
of the topological scattering in the context of quantum mechanics on a cone
\cite{5}, and the investigations on the interaction of a quantum
system with conical singularities\cite{6b,gv} and on quantum mechanics on
topological defects spacetimes
\cite{6}.

Therefore, taking into account that we have to consider the topology of
spacetime in order to describe completely a given physical system, we want
to address the question of how the nontrivial topology could affect the
energy levels by shifting the atomic spectral lines. For the purpose of
investigating this problem, a calculation of the energy levels shifts of the
hydrogen atom is carried out in the spacetimes of an infinitely thin cosmic
string\cite{7} and of a point-like global monopole\cite{8}.

Topological defects may arise in gauge models with spontaneous symmetry
breaking. They can be of various types such as monopoles, domain walls,
cosmic strings and their hybrids\cite{9}. They may have been formed during
universe expansion and their nature depends on the topology of the vacuum
manifold of the theory under consideration\cite{kb}. The richness of
the new ideas they brought along to general relativity seems to justify
the interest in the study of these structures, and specifically the role
played by their topological features at the atomic level.

The gravitational field of a cosmic string is quite remarkable; a particle
placed at rest around a straight, infinite, static cosmic string will not be
attracted to it; there is no local gravity. The spacetime around a cosmic
string is locally flat but not globally. The external gravitational field
due to a cosmic string may be approximately described by a commonly called
conical geometry. The nontrivial topology of this spacetime leads to a
number of interesting effects like, for example, gravitational lensing \cite
{10}, emission of radiation by a freely moving particle\cite{11},
electrostatic self-force\cite{12} on an electric charge at rest and the
so-called gravitational Aharonov-Bohm effect\cite{14} among other.

The spacetime of a point-like global monopole has also some unusual
properties. It possesses a deficit solid angle $\Delta =32\pi ^{2}G\eta ^{2}$%
, $\eta $ being the energy scale of symmetry breaking. Test particles in
this spacetime experiences a topological scattering by an angle $\pi \Delta
/2$ irrespective of their velocity and their impact parameter. Also in this
case, the nontrivial topology of spacetime, as well as its curvature, which
are due to the deficit solid angle leads to a number of interesting
 effects\cite{gv1,pec}
which are not present in flat Minkowski spacetime.

In this paper, we deal with the interesting problem concerning the
 modifications of the
energy levels of a hydrogen atom placed in the gravitational fields of a
cosmic string and of a global monopole. In order to investigate this
 problem further we determine the
solutions of the corresponding Dirac equations and the energy
levels of a hydrogen atom under the influence of these gravitational fields.
To do these calculations we shall make the following
assumptions: $(i)$ The atomic nucleus is not affected by
the presence of the defect. $(ii)$ The atomic nucleus is located on the
defect. With these, to do our calculations accordingly would have been
possible and doing so it affords an explicit demonstration of the effects of
spacetime topology on the shifts in the atomic spectral lines of the
hydrogen atom.

A similar problem concerning the effects of gravitational fields at atomic
level has been considered before. As example of some works on this topic, we
can mention [4-8] which obtained the expressions for the shifts in the
energy levels of an atom caused by its interaction with the curvature of
spacetime and also a recent paper\cite{b} which calculated the atomic energy
level shifts of atoms placed in strong gravitational fields near collapsing
spheroidal masses.

The results obtained in this paper are related to the previous ones[4-8]
connected with this topic, in the sense that we also study the effect of
gravitational fields at the atomic level, however, our calculation provides
an interesting new example of an effect at atomic scale which can be thought
of as a consequence of the nontrivial topology of spacetime
 and this aspect was not taken into account by previous
works[4,8].

In the case of an infinitely thin cosmic string spacetime, the shifts in the
energy levels depend on the angle deficit and for the global monopole
spacetime these shifts depend on the deficit solid angle. In both situations
these effects vanish when these angle deficits
vanish, as it should be.

This paper is organized as follows. In section II we obtain the solution of
the Dirac equation and we calculate the energy shifts experienced by a
hydrogen atom placed in the gravitational field of a cosmic string. In
section III we also obtain the solutions of the Dirac equation and we
calculate the modifications of the spectrum of a
 hydrogen atom in the gravitational
field of a global monopole. Finally, in section IV, we draw some conclusions.

\vskip 1.0 cm

{\bf II. Relativistic hydrogen atom in the spacetime of the cosmic string }

In what follows we will study the behaviour of  a hydrogen
atom in the spacetime of a cosmic
string. The line element
corresponding to the cosmic string spacetime\cite{7} is given, in
spherical coordinates, by
\begin{equation}
ds^{2}=-c^{2}dt^{2}+dr^{2}+r^{2}d\theta ^{2}+\alpha ^{2}r^{2}\sin ^{2}\theta
d\phi ^{2}.  \label{100}
\end{equation}

The parameter $\alpha =1-\frac{4G}{c^{2}}\bar{\mu}$ runs in the interval $%
(0,1]$, with $\bar{\mu}$ being the linear mass density of the cosmic string.

Let us consider the generally covariant form of the Dirac equation
 which is given by

\begin{equation}
\left[ i\gamma ^{\mu }\left( x\right) \left( \partial _{\mu }+\Gamma _{\mu
}\left( x\right) +i\frac{eA_{\mu }}{\hbar c}\right) -\frac{\mu c}{\hbar }%
\right] \Psi \left( x\right) =0,  \label{103}
\end{equation}
where $\mu $ is the mass of the particle, $A_{\mu }$ is an external
electromagnetic potential and $\Gamma _{\mu }\left( x\right) $ are the
spinor affine connections which can be expressed in terms of the set of
tetrad fields $e^{\mu}_{(a)}(x)$ and the standard flat
 spacetime $\gamma^{(a)}$ Dirac matrices as

\begin{eqnarray}
\Gamma_{\mu}=\frac{1}{4}\gamma^{(a)}\gamma^{(b)}e_{(a)}^{\nu}
(\partial_{\mu}e_{(b)\nu} - \Gamma^{\sigma}_{\mu \nu}e_{(b)\sigma}).
\label{fim}
 \end{eqnarray}

The generalized Dirac matrices $\gamma ^{\mu
}\left( x\right) $ satisfies the anticommutation relations
\[
\left\{ \gamma ^{\mu }\left( x\right) ,\gamma ^{\nu }\left( x\right)
\right\} =2g^{\mu \nu }\left( x\right) ,
\]
and are defined by
\begin{equation}
\gamma ^{\mu }\left( x\right) =e_{\left( a\right) }^{\mu }\left( x\right)
\gamma ^{\left( a\right) },  \label{104}
\end{equation}
where $e_{\left( a\right) }^{\mu }\left( x\right) $ obeys the
relation $\eta ^{ab}e_{\left( a\right) }^{\mu }\left( x\right) e_{\left(
b\right) }^{\nu }\left( x\right) =g^{\mu \nu };$ $\mu ,\,\nu =0,1,2,3$ are
tensor indices and $a,\,b=0,1,2,3$ are tetrad indices.

In this paper, the following explicit forms of the constant Dirac matrices
will be taken
\begin{equation}
\gamma ^{\left( 0\right) }=\left(
\begin{array}{cc}
{\bf 1} & 0 \\
0 & {\bf -1}
\end{array}
\right) ;\text{ }\gamma ^{\left( i\right) }=\left(
\begin{array}{cc}
0 & \sigma ^{i} \\
-\sigma ^{i} & 0
\end{array}
\right) ;\text{ }i=1,2,3\,, \label{105}
\end{equation}
where $\sigma ^{i}$ are the usual Pauli matrices.

In order to write the Dirac equation in this spacetime, let us take the
tetrads $e_{\left( a\right) }^{\mu }(x)$ as
\begin{equation}
e_{\left( a\right) }^{\mu }(x)=\left(
\begin{array}{cccc}
1 & 0 & 0 & 0 \\
0 & \sin \theta \cos \phi & \sin \theta \sin \phi & \cos \theta \\
0 & \frac{\cos \theta \cos \phi }{r} & \frac{\cos \theta \sin \phi }{r} & -%
\frac{\sin \theta }{r} \\
0 & -\frac{\sin \phi }{\alpha r\sin \theta } & \frac{\cos \phi }{\alpha
r\sin \theta } & 0
\end{array}
\right) .  \label{101}
\end{equation}

Thus using (\ref{101}), we obtain the following expressions for the
generalized Dirac matrices $\gamma ^{\mu }\left( x\right) $
\begin{eqnarray}
\gamma ^{0}\left( x\right) &=&\gamma ^{\left( 0\right) },  \nonumber \\
\gamma ^{1}\left( x\right) &=&\gamma ^{\left( r\right) },  \nonumber \\
\gamma ^{2}\left( x\right) &=&\frac{\gamma ^{\left( \theta \right) }}{r},
\nonumber \\
\gamma ^{3}\left( x\right) &=&\frac{\gamma ^{\left( \phi \right) }}{\alpha
r\sin \theta },  \label{114}
\end{eqnarray}
where
\begin{equation}
\left(
\begin{array}{c}
\gamma ^{\left( r\right) } \\
\gamma ^{\left( \theta \right) } \\
\gamma ^{\left( \phi \right) }
\end{array}
\right) =\left(
\begin{array}{ccc}
\cos \phi \sin \theta & \sin \phi \sin \theta & \cos \phi \\
\cos \phi \cos \theta & \sin \phi \cos \theta & -\sin \phi \\
-\sin \phi & \cos \phi & 0
\end{array}
\right) \left(
\begin{array}{c}
\gamma ^{\left( 1\right) } \\
\gamma ^{\left( 2\right) } \\
\gamma ^{\left( 3\right) }
\end{array}
\right) .  \label{110}
\end{equation}

The covariant Dirac Eq. (\ref{103}), written in the spacetime of a
cosmic string is then given by

\begin{eqnarray}
&&\left[ i\hbar
\mathop{\textstyle\sum}%
\nolimits^{r}\partial _{r}+i\hbar \frac{%
\mathop{\textstyle\sum}%
{}^{\theta }}{r}\partial _{\theta }+i\hbar \frac{%
\mathop{\textstyle\sum}%
{}^{\phi }}{\alpha r\sin \theta }\partial _{\phi }\right.
\nonumber \\
&&\left. +i\hbar \frac{1}{2r}\left( 1-\frac{1}{\alpha }\right) \left(
\mathop{\textstyle\sum}%
\nolimits^{r}+\cot \theta
\mathop{\textstyle\sum}%
\nolimits^{\theta }\right) -\frac{eA_{0}}{c}-\gamma ^{\left( 0\right) }
\mu c +\frac{E}{c}%
\right] \chi \left( \vec{r}\right)  =0, \label{119}
\end{eqnarray}
where $\sum^{r}$, $\sum^{\theta }$ and $\sum^{\phi }$ are defined by

\begin{equation}
\mathop{\textstyle\sum}%
\nolimits^{r}\equiv \gamma ^{\left( 0\right) }\gamma ^{\left( r\right) };%
\text{ }%
\mathop{\textstyle\sum}%
\nolimits^{\theta }\equiv \gamma ^{\left( 0\right) }\gamma ^{\left( \theta
\right) };\text{ }%
\mathop{\textstyle\sum}%
{}^{\phi }\equiv \gamma ^{\left( 0\right) }\gamma ^{\left( \phi \right) },
\label{118}
\end{equation}
and we have chosen $\Psi \left( x\right) $ as

\begin{equation}
\Psi \left( x\right) =e^{-i\frac{E}{\hbar }t}\chi \left( \vec{r}\right) ,
\label{116}
\end{equation}
which comes from the fact that the spacetime under consideration is static.

We must now turn our attention to the solution of the equation
 for $\chi (\vec{r}) $. Then,
let us assume that the solutions of Eq. (\ref{119}) are of the form
\begin{equation}
\chi \left( \vec{r}\right) =r^{-\frac{1}{2}\left( 1-\frac{1}{\alpha }\right)
}\left( \sin \theta \right) ^{-\frac{1}{2}\left( 1-\frac{1}{\alpha }\right)
}R\left( r\right) \Theta \left(
\theta \right) \Phi \left( \phi \right) .  \label{120}
\end{equation}

Thus, substituting   Eq. (\ref{120}) into
 (\ref{119}), we obtain the following radial equation
\begin{equation}
\left( c%
\mathop{\textstyle\sum}%
\nolimits_{r}^{\prime }p_{r}+i\hbar c\frac{%
\mathop{\textstyle\sum}%
{}_{r}^{\prime }}{r} \gamma ^{\left( 0\right) }k_{\left( \alpha \right)
}+eA_{0}+\mu c^{2}\gamma ^{\left( 0\right) }\right) R\left( r\right)
=ER\left( r\right) .  \label{136}
\end{equation}
where
\begin{equation}
k_{\left( \alpha \right) }=\pm \left( j_{\left( \alpha \right) }+\frac{1}{2}%
\right) =\pm \left[ j+\frac{1}{2}+m\left( \frac{1}{\alpha }-1\right) \right]
\label{135}
\end{equation}
are the eigenvalues of the generalized spin-orbit operator $K_{\left(
\alpha \right) }$ in the spacetime of a cosmic string and $j_{(\alpha)} $
corresponds to the eigenvalues of the generalized total angular momentum
operator. The
operator $K_{\alpha} $ is given by

\begin{equation}
\hbar \gamma ^{\left( 0\right) } K_{\left( \alpha \right) }=\vec{%
\mathop{\textstyle\sum}%
}\cdot \vec{L}_{\left( \alpha \right) }+\hbar ,  \label{128b}
\end{equation}
with $\vec{\Sigma}= (\Sigma^{r},\;\Sigma^{\theta},\;\Sigma^{\phi}) $
and  $\vec{L}_{\left( \alpha \right) }$ is the generalized angular
momentum operator\cite{gv} in the spacetime of the cosmic string, which is
such that $\vec{L}_{\left( \alpha \right) }^{2}Y_{l_{\left( \alpha \right)
}}^{m_{\left( \alpha \right) }}\left( \theta ,\phi \right) =\hbar
^{2}l_{\left( \alpha \right) }\left( l_{\left( \alpha \right) }+1\right) ,$
with $Y_{l_{\left( \alpha \right) }}^{m_{\left( \alpha \right) }}\left(
\theta ,\phi \right) $ being the generalized spherical harmonics in the sense
that $m_{\left( \alpha \right) }$ and $l_{\left( \alpha \right) }$ are not
necessarily integers. The parameters $m_{\left( \alpha \right) }$ and $%
l_{\left( \alpha \right) }$ are given, respectively, by $m_{\left( \alpha
\right) }\equiv \frac{m}{\alpha }$ and $l_{\left( \alpha \right) }\equiv
n+m_{\left( \alpha \right) }=l+m\left( \frac{1}{\alpha }-1\right) ,$ $%
l=0,1,2,...$ $n-1$, $l$ is the orbital angular momentum quantum number, $m$
is the magnetic quantum number and $n$ is the principal quantum number.

Let us choose the following two-dimensional representation for $%
\mathop{\textstyle\sum}%
\nolimits_{r}^{\prime }$ and $\gamma ^{\left( 0\right) }$%
\begin{equation}
\mathop{\textstyle\sum}%
\nolimits_{r}^{\prime }\equiv \left(
\begin{array}{cc}
0 & -i \\
i & 0
\end{array}
\right) ;\text{ }\gamma ^{\left( 0\right) }\equiv \left(
\begin{array}{cc}
1 & 0 \\
0 & -1
\end{array}
\right) .  \label{137}
\end{equation}
Now, let us assume that the radial solution can be written as
\begin{equation}
R(r)=\frac{1}{r}\left( \newline
\begin{array}{c}
-iF\left( r\right)  \\
G\left( r\right)
\end{array}
\right) .  \label{138}
\end{equation}
Then,  Eq. (\ref{136}) decomposes into the coupled equations

\begin{equation}
-i\left( \hbar c\right) ^{-1}\left[ E-E_{0}+\frac{e^{2}}{r}\right] F(r)+%
\frac{dG(r)}{dr}+\frac{k_{(\alpha )}}{r}G(r)=0,  \label{144}
\end{equation}
and
\begin{equation}
-i\left( \hbar c\right) ^{-1}\left[ E+E_{0}+\frac{e^{2}}{r}\right] G(r)+%
\frac{dF(r)}{dr}-\frac{k_{(\alpha )}}{r}F(r)=0,  \label{145}
\end{equation}
where $E_{0}=\mu c^2 $ is the rest energy of the electron. Note
that in obtaining these equations use was made  of the fact that $A_{0}=-e/r$.

The solutions of these equations are given in terms of the confluent
 hypergeometric
 function $ M(a,b;x)$ as

\begin{eqnarray}
F(r) &=&-i\sqrt{\frac{Q}{T}}\frac{e^{-rD}}{2}\left( rD\right) ^{\gamma
_{\left( \alpha \right) }-1}\left[ M\left( \gamma _{\left( \alpha \right)
}-1+\tilde{P},2\gamma _{\left( \alpha \right) }-1;2rD\right) \right.
\nonumber \\
&&\left. +\frac{\left( \gamma _{\left( \alpha \right) }-1+\tilde{P}\right) }{%
\left( k_{\left( \alpha \right) }+\tilde{Q}\right) }M\left( \gamma _{\left(
\alpha \right) }+\tilde{P},2\gamma _{\left( \alpha \right) }-1;2rD\right) %
\right] ,  \label{181}
\end{eqnarray}
and
\begin{eqnarray}
G(r) &=&\frac{e^{-rD}}{2}\left( rD\right) ^{\gamma _{\left( \alpha \right)
}-1}\left[ M\left( \gamma _{\left( \alpha \right) }-1+\tilde{P},2\gamma
_{\left( \alpha \right) }-1;2rD\right) \right.   \nonumber \\
&&\left. -\frac{\left( \gamma _{\left( \alpha \right) }-1+\tilde{P}\right) }{%
\left( k_{\left( \alpha \right) }+\tilde{Q}\right) }M\left( \gamma _{\left(
\alpha \right) }+\tilde{P},2\gamma _{\left( \alpha \right) }-1;2rD\right) %
\right] ,  \label{180}
\end{eqnarray}
where $T=\frac{E_{0}-E}{\hbar c};$ $Q=\frac{E_{0}+E}{\hbar c},$ $D=\sqrt{TQ}=%
\frac{\sqrt{E_{0}^{2}-E^{2}}}{\hbar c};$ $\gamma _{\left( \alpha \right) }=1+%
\sqrt{k_{\left( \alpha \right) }^{2}-\tilde{\alpha}^{2}};$ $\ \tilde{P}%
\equiv \frac{\tilde{\alpha}}{2}\left( \sqrt{T/Q}-\sqrt{Q/T}\right) ;$ $%
\tilde{Q}\equiv \frac{\tilde{\alpha}}{2}\left( \sqrt{T/Q}+\sqrt{Q/T}\right) ,
$ with $\tilde{\alpha}=\frac{e^{2}}{\hbar c}\approx\frac{1}{137}$
 being the fine
structure constant.

The solutions given by (\ref{181}) and (\ref{180}) are divergent, unless the
following condition is fulfilled
\begin{equation}
\gamma _{\left( \alpha \right) }-1+\tilde{P}=-n;\text{ }n =0,1,2...,
\label{b}
\end{equation}
which means that
\begin{equation}
\frac{1}{2}\tilde{\alpha}\left( \sqrt{\frac{T}{Q}}-\sqrt{\frac{Q}{T}}\right)
=-\left( n +\gamma _{\left( \alpha \right) }-1\right) .
  \label{c}
\end{equation}
From this equation we may infer that the energy
eigenvalues are given by
\begin{equation}
E=E_{0}\left[ 1+\tilde{\alpha}^{2}\left( n +\left| k_{\left( \alpha
\right) }\right| \sqrt{1-\tilde{\alpha}^{2}k_{\left( \alpha \right) }^{-2}}%
\right) ^{-2}\right] ^{-\frac{1}{2}}.  \label{184}
\end{equation}
This equation exhibits the angle deficit dependence of the energy levels. It
is helpful to introduce the quantum number $n_{(\alpha)}$ that corresponds
 to the
principal quantum
number of the nonrelativistic theory when $ \alpha = 1$,
\begin{equation}
n_{(\alpha)}=n+j_{\left( \alpha \right) }+\frac{1}{2}.  \label{185}
\end{equation}
Therefore, Eq. (\ref{184}) may be cast in the form
\begin{equation}
E_{n_{(\alpha)},j_{\left( \alpha \right) }}=E_{0}\left\{ 1+
\tilde{\alpha}^{2}\left[\left( n_{(\alpha)}-j_{\left( \alpha \right) }-
\frac{1}{2}\right) + \left( j_{\left(
\alpha \right) }+\frac{1}{2}\right) \sqrt{1-\tilde{\alpha}^{2}\left(
j_{\left( \alpha \right) }+\frac{1}{2}\right) ^{-2}}\right] ^{-2}\right\} ^{-%
\frac{1}{2}}.  \label{186}
\end{equation}

This equation can be written in a way which is better
 suited to physical interpretation. Thus,
as $\tilde{\alpha}\ll 1,$ we can expand  Eq. (\ref{186}) in a powers
 of $\tilde{%
\alpha}$, and as a result we get the following leading terms
\begin{equation}
E_{n_{(\alpha)},j_{\left( \alpha \right) }}=E_{0}-
E_{0}\frac{\tilde{\alpha}^{2}}{2n_{(\alpha)}^{2}}%
+E_{0}\frac{\tilde{\alpha}^{4}}{2n_{(\alpha)}^{4}}
\left( \frac{3}{4}-\frac{n_{(\alpha)}}{j_{\left(
\alpha \right) }+\frac{1}{2}}\right) \text{.}  \label{190}
\end{equation}

The first term corresponds to the rest energy of the electron; the second
one gives the energy of the bound states in the non-relativistic
approximation and the third one corresponds to the relativistic correction.
 Note that these last two terms depend on the deficit angle.
The further terms can be neglected in comparison with these first three
terms.

Now, let us consider the total shift in the energy between the states with
$j=n-\frac{1}{2}$, and $%
j=\frac{1}{2},$ for a given $n$. This shift is given by
\begin{eqnarray}
\Delta E_{n_{(\alpha)},j_{\left( \alpha \right) }}
&=&\frac{\mu e^{8}}{\hbar ^{4}c^{2}n_{(\alpha)}^{3}}\left( \frac{n_{(\alpha)}-
1}{2\left[ n_{(\alpha)}+m\left(
\frac{1}{\alpha }-1\right) \right] \left[ 1+m\left( \frac{1}{\alpha }%
-1\right) \right] }\right) .  \label{191}
\end{eqnarray}
One important characteristic of  Eq. (\ref{186}) is that it contains a
dependence on $n$, $j$ and $\alpha $.  The dependence on
$ \alpha $ corresponds to an analogue of the
electromagnetic Aharonov-Bohm effect for bound states, but now in the
gravitational context. Therefore, the interaction with the topology(conical
singularity) causes the energy levels to change. Note that the presence of
the cosmic string destroys the degeneracy of all the levels, corresponding
to $l=0$ and $l=1$, and destroys partially this degeneracy for the other
sublevels. Therefore, as the occurrence of degeneracy can often be ascribed
to some symmetry property of the physical system, the fact that the presence
of the cosmic string destroys the degeneracy means that there is a break of
the original symmetry. Observe that for $\alpha =1$, the results reduce to
the flat Minkowski spacetime case as expected.

As a estimation of the effect of the cosmic string on the energy shift of
the hydrogen atom, let us consider $\alpha =1-10^{-6}$ which corresponds to
GUT cosmic strings. Using this value into  Eq. (\ref{191}), we conclude
that the presence of the cosmic string reduces the energy of the level of
the states $2P_{1/2}(n=2$, $l=1$, $j=l-\frac{1}{2}=\frac{1}{2}$, $m=1)$ to
about $10^{-4}\%$ in comparison with the  flat spacetime value. This
decrease is of the order of the measurable Zeeman effect in carbon
atoms for $2P$ states when submitted, for example, to an external
magnetic field with strength to about tens of Tesla. Therefore, this shift
in energy levels produced by a cosmic string is measurable as well.

Finally, we can write down the general solution to  Eq. (\ref{103})
corresponding to a hydrogen atom placed in the background spacetime of a
cosmic string. Thus, it reads
\begin{eqnarray}
\Psi _{l_{\left( \alpha \right) },j_{\left( \alpha \right) }=l_{\left(
\alpha \right) }+\frac{1}{2},m_{\left( \alpha \right) }}\left( x\right)
&=&e^{-i\frac{Et}{\hbar }}r^{-\frac{1}{2}\left( 1-\frac{1}{\alpha }\right)
}\left( \sin \theta \right) ^{-\frac{1}{2}\left( 1-\frac{1}{\alpha }\right) }
\nonumber \\
&&\times F_{\left( \alpha \right) }\left( r\right) \left(
\begin{array}{c}
\sqrt{\frac{l_{\left( \alpha \right) }+m_{\left( \alpha \right) }+\frac{1}{2}%
}{2l_{\left( \alpha \right) }+1}}Y_{l\left( \alpha \right) }^{m_{\left(
\alpha \right) }-\frac{1}{2}}\left( \theta ,\phi \right)  \\
\sqrt{\frac{l_{\left( \alpha \right) }-m_{\left( \alpha \right) }+\frac{1}{2}%
}{2l_{\left( \alpha \right) }+1}}Y_{l\left( \alpha \right) }^{m_{\left(
\alpha \right) }+\frac{1}{2}}\left( \theta ,\phi \right)
\end{array}
\right) ,  \label{195}
\end{eqnarray}
and
\begin{eqnarray}
\Psi _{l_{\left( \alpha \right) },j_{\left( \alpha \right) }=l_{\left(
\alpha \right) }-\frac{1}{2},m_{\left( \alpha \right) }}\left( x\right)
&=&e^{-i\frac{Et}{\hbar }}r^{-\frac{1}{2}\left( 1-\frac{1}{\alpha }\right)
}\left( \sin \theta \right) ^{-\frac{1}{2}\left( 1-\frac{1}{\alpha }\right) }
\nonumber \\
&&\times G_{\left( \alpha \right) }\left( r\right) \left(
\begin{array}{c}
-\sqrt{\frac{l_{\left( \alpha \right) }-m_{\left( \alpha \right) }+\frac{1}{2%
}}{2l_{\left( \alpha \right) }+1}}Y_{l\left( \alpha \right) }^{m_{\left(
\alpha \right) }-\frac{1}{2}}\left( \theta ,\phi \right)  \\
\sqrt{\frac{l_{\left( \alpha \right) }+m_{\left( \alpha \right) }+\frac{1}{2}%
}{2l_{\left( \alpha \right) }+1}}Y_{l\left( \alpha \right) }^{m_{\left(
\alpha \right) }+\frac{1}{2}}\left( \theta ,\phi \right)
\end{array}
\right) ,  \label{196}
\end{eqnarray}
where $F_{\left( \alpha \right) }\left( r\right) $ and $G_{\left( \alpha
\right) }\left( r\right) $ are given by  Eqs. (\ref{181}) and (\ref{180}%
), respectively, and the index $ \alpha$ was introduced to emphasize
 the dependence of these functions on this parameter.

Note that the solutions depend on the topological features of the
 spacetime of a cosmic string whose influence appears codified
 in the parameter $\alpha $
associated with the presence of the cosmic string and this
 is the point at issue here.

\newpage

{\bf III.} {\bf Relativistic hydrogen atom in the presence of a global
monopole}

In continuation of the preceding consideration, in this section we shall be
concerned with the study of the influence of a global monopole on the states
of a hydrogen atom.

 The solution corresponding to a global monopole in a $O\left( 3\right) $
broken symmetry model has been investigated by Barriola and Vilenkin\cite{8}.

Far away from the global monopole core we can neglect the mass term and as a
consequence the main effects are produced by the solid deficit angle. The
respective metric in the Einstein theory of gravity can be written as\cite{8}

\begin{equation}
ds^{2}=-c^{2}dt^{2}+dr^{2}+b^{2}r^{2}\left( d\theta ^{2}+\sin ^{2}\theta
d\phi ^{2}\right) ,  \label{200}
\end{equation}
where $b^{2}=1-8\pi G\eta ^{2}$, the parameter $\eta $ being the energy
scale of symmetry breaking.

Now, let us choose the tetrad as

\begin{equation}
e_{\left( a\right) }^{\mu }=\left(
\begin{array}{cccc}
1 & 0 & 0 & 0 \\
0 & \sin \theta \cos \phi & \sin \theta \sin \phi & \cos \theta \\
0 & \frac{\cos \theta \cos \phi }{br} & \frac{\cos \theta \sin \phi }{br} & -%
\frac{\sin \theta }{br} \\
0 & -\frac{\sin \phi }{br\sin \theta } & \frac{\cos \phi }{br\sin \theta } &
0
\end{array}
\right) .  \label{201}
\end{equation}
Therefore, the generalized and flat spacetime Dirac matrices are related by

\begin{eqnarray}
\gamma ^{0}\left( x\right) &=&\gamma ^{\left( 0\right) },  \nonumber \\
\gamma ^{1}\left( x\right) &=&\gamma ^{\left( r\right) },  \nonumber \\
\gamma ^{2}\left( x\right) &=&\frac{\gamma ^{\left( \theta \right) }}{br},
\nonumber \\
\gamma ^{3}\left( x\right) &=&\frac{\gamma ^{\left( \phi \right) }}{br\sin
\theta }.  \label{214}
\end{eqnarray}
where, $\gamma ^{\left( r\right) }$, $\gamma ^{\left( \theta \right) }$and $%
\gamma ^{\left( \phi \right) }$ were defined in the previous section.

Proceeding in analogy with section II we find that the generalized Dirac
equation can be written, in this background spacetime, as
\begin{eqnarray}
&&\left[ i\hbar
\mathop{\textstyle\sum}%
\nolimits^{r}\partial _{r}+i\hbar \frac{%
\mathop{\textstyle\sum}%
{}^{\theta }}{br}\partial _{\theta }+i\hbar \frac{%
\mathop{\textstyle\sum}%
{}^{\phi }}{br\sin \theta }\partial _{\phi }\right.   \nonumber
\\
&&\left. +i\hbar \frac{1}{r}\left( 1-\frac{1}{b}\right)
(\mathop{\textstyle\sum}%
\nolimits^{r} + \cot \theta\mathop{\textstyle\sum}%
\nolimits^{\theta} )+\frac{e^{2}}{rc}-\gamma ^{\left( 0\right) }\mu c +
\frac{E}{c}
\right] \chi
\left( \vec{r}\right) \left. =0\right. ,  \label{219}
\end{eqnarray}
where Eq. (\ref{116}) has been used in obtaining the above result.

Now, let us assume that the solution of  Eq. (\ref{219}) can be written as
\begin{equation}
\chi \left( \vec{r}\right) =r^{-\left( 1-\frac{1}{b}\right) }R
\left( r\right) \Theta \left(
\theta \right) \Phi \left( \phi \right) ,  \label{220}
\end{equation}

Using  Eq. (\ref{220}),  Eq.  (\ref{219}) turns into the simple form

\begin{equation}
\left( c%
\mathop{\textstyle\sum}%
\nolimits_{r}^{\prime }p_{r}+i\hbar c\frac{%
\mathop{\textstyle\sum}%
{}_{r}^{\prime }}{r} \gamma ^{\left( 0\right) }k_{\left( b\right) }-%
\frac{e^{2}}{r}+\mu c^{2}\gamma ^{\left( 0\right) }\right) R\left( r\right)
=ER\left( r\right) ,  \label{221}
\end{equation}
where

\begin{eqnarray}
k_{\left( b\right) } &=&\pm \left( \frac{j^{2}}{b^{2}}+\frac{j}{b^{2}}+\frac{%
1}{4}\right) ^{\frac{1}{2}}  \nonumber \\
&=&\pm \left[ \left( \frac{j}{b}+\frac{1}{2}\right) ^{2}+\frac{j}{b}\left(
\frac{1}{b}-1\right) \right] ^{\frac{1}{2}}  \label{225}
\end{eqnarray}
are the eigenvalues of the generalized spin-orbit operator $K_{\left(
b\right) }$ in the spacetime of a global monopole which is given by

\[
\hbar K_{\left( b\right) }=\gamma ^{\left( 0\right) }\left[ \vec{%
\mathop{\textstyle\sum}%
}^{\prime \prime }\cdot \vec{L}_{\left( b\right) }+\hbar \right] ,
\]
where $\sum^{\prime \prime }=\left(
\begin{array}{cc}
\vec{\sigma} & 0 \\
0 & \vec{\sigma}
\end{array}
\right) .$

In the present case the generalized angular momentum will be denoted by $%
L_{\left( b\right) }$. It is such that $\vec{L}_{\left( b\right)
}^{2}Y_{l}^{m}\left( \theta ,\phi \right) =\frac{\hbar ^{2}}{b^{2}}l\left(
l+1\right) ,$ $l=0,1,2,...$ $n-1$, and $\vec{L}_{\left( b\right) }=\frac{%
\vec{L}}{b}$ is the angular momentum in the spacetime of a global monopole
\cite{gv1}. Using the same procedure as in the previous section, we find
\begin{eqnarray}
F_{\left( b\right) }(r) &=&-i\sqrt{\frac{Q}{T}}\frac{e^{-rD}}{2}\left(
rD\right) ^{\gamma _{\left( b\right) }-1}\left[ M\left( \gamma _{\left(
b\right) }-1+\tilde{P},2\gamma _{\left( b\right) }-1;2rD\right) \right.
\nonumber \\
&&\left. +\frac{\left( \gamma _{\left( b\right) }-1+\tilde{P}\right) }{%
\left( k_{\left( b\right) }+\tilde{Q}\right) }M\left( \gamma _{\left(
b\right) }+\tilde{P},2\gamma _{\left( b\right) }-1;2rD\right) \right] ,
\label{223}
\end{eqnarray}
and
\begin{eqnarray}
G_{\left( b\right) }(r) &=&\frac{e^{-rD}}{2}\left( rD\right) ^{\gamma
_{\left( b\right) }-1}\left[ M\left( \gamma _{\left( b\right) }-1+\tilde{P}%
,2\gamma _{\left( b\right) }-1;2rD\right) \right.   \nonumber \\
&&\left. -\frac{\left( \gamma _{\left( b\right) }-1+\tilde{P}\right) }{%
\left( k_{\left( b\right) }+\tilde{Q}\right) }M\left( \gamma _{\left(
b\right) }+\tilde{P},2\gamma _{\left( b\right) }-1;2rD\right) \right] ,
\label{222}
\end{eqnarray}
where $\gamma _{\left( b\right) }=1+\sqrt{k_{\left( b\right) }^{2}-\tilde{%
\alpha}^{2}\text{ }}\,$; $T,$ $Q,$ $M,$ $\tilde{P}$ and $\tilde{Q}$ are the
same defined previously. The index $b$ in the functions $F$ and $G$ indicates
their dependence on this parameter. These functions are, formally, the same
used in the previous section.

By the use of condition (\ref{b}) with the interchange of
$ \gamma_{(\alpha)}$ by $ \gamma_{(b)}$, we obtain
the following spectrum of energy eigenvalues
\begin{equation}
E_{n_{(b)},j_{\left( b\right) }}=E_{0}\left\{ 1+\tilde{\alpha}^{2}
\left[ n_{(b)}-\left| k_{\left( b\right) }\right| +
\left| k_{\left( b\right) }\right| \sqrt{1-%
\tilde{\alpha}^{2}k_{\left( b\right) }^{-2}}\right] ^{-2}\right\} ^{-\frac{1%
}{2}}.  \label{230}
\end{equation}
in which we have defined $n_{(b)}$ as a number which reduces to the principal
quantum number when $b=1$ and is given by
\begin{equation}
n_{(b)}=n+\left| k_{\left( b\right) }\right| =n+\left[ \left( \frac{j}{b}+%
\frac{1}{2}\right) ^{2}+\frac{j}{b}\left( \frac{1}{b}-1\right) \right] ^{%
\frac{1}{2}}.  \label{229}
\end{equation}

Then, expanding Eq. (\ref{230}) in a series of powers of $\tilde{\alpha%
}$, we have the following leading terms
\begin{eqnarray}
E_{n_{(b)},j_{\left( b\right) }} &=&E_{0}-
E_{0}\frac{\tilde{\alpha}^{2}}{2n_{(b)}^{2}}
+E_{0}\frac{\tilde{\alpha}^{4}}{2n_{(b)}^{4}}\left( \frac{3}{4}-
\frac{n_{(b)}}{\left[
\left( \frac{j}{b}+\frac{1}{2}\right) ^{2}+\frac{j}{b}\left( \frac{1}{b}%
-1\right) \right] ^{\frac{1}{2}}}\right) \text{,}  \label{234}
\end{eqnarray}
which tell us what the dependence of each term with the parameter $ b$ is.
In this case, the shift in the energy between the energy levels with
$j=n-\frac{1}{2}$ and $j=
\frac{1}{2},$ for a given $n$, is

\begin{eqnarray}
\Delta E_{n_{(b)},j_{\left( b\right) }} &=&\frac{\mu e^{8}}
{2\hbar ^{4}c^{2}n^{3}}
\nonumber \\
&&\left\{ \frac{\left[ \left( \frac{n_{(b)}}{b}-\frac{1}{2}
\left( \frac{1}{b}%
-1\right) \right) ^{2}+\left( \frac{n_{(b)}}{b}-\frac{1}{2b}
\right) \left( \frac{1%
}{b}-1\right) \right] ^{\frac{1}{2}}-\left[ \left( \frac{1}{2b}+\frac{1}{2}%
\right) ^{2}+\frac{1}{2b}\left( \frac{1}{b}-1\right) \right] ^{\frac{1}{2}}}{%
\left[ \left( \frac{1}{2b}+\frac{1}{2}\right) ^{2}+\frac{1}{2b}\left( \frac{1%
}{b}-1\right) \right] ^{\frac{1}{2}}\left[ \left( \frac{n_{(b)}}
{b}-\frac{1}{2}%
\left( \frac{1}{b}-1\right) \right) ^{2}+\left( \frac{n_{(b)}}
{b}-\frac{1}{2b}%
\right) \left( \frac{1}{b}-1\right) \right] ^{\frac{1}{2}}}\right\} .
\label{191}
\end{eqnarray}
This equation reduces to the same result of the flat spacetime in the
absence of the global monopole ($b=1$).

It is worth noticing from Eq. (\ref{234}) that the presence of the
monopole does not break the degeneracy of the energy levels and as in the
case of a cosmic string.

As an estimation of the shift in the energy levels, let us consider a grand
unified (GUT) monopole in which $b^{2}=1-10^{-6}$. Using this value into
Eq. (\ref{234}) we conclude that the presence of the monopole reduces
the relativistic correction of the energy of the level $2P_{1/2}(n=2$, $l=1$%
, $j=l-\frac{1}{2}=\frac{1}{2}$, $m=1)$ in approximately $10^{-4}\%$ as
compared with the result of the flat Minkowski spacetime.

Finally, let us write down the general solution for this case. It reads as
\begin{eqnarray}
\Psi _{l,j=l+\frac{1}{2},m}\left( x\right)  &=&e^{-i\frac{Et}{\hbar }%
}r^{-\left( 1-\frac{1}{b}\right) }  \nonumber \\
&&\times F_{\left( b\right) }\left( r\right) \left(
\begin{array}{c}
\sqrt{\frac{l+m+\frac{1}{2}}{2l+1}}Y_{l}^{m-\frac{1}{2}}\left( \theta ,\phi
\right)  \\
\sqrt{\frac{l-m+\frac{1}{2}}{2l+1}}Y_{l}^{m+\frac{1}{2}}\left( \theta ,\phi
\right)
\end{array}
\right) ,  \label{240}
\end{eqnarray}
and
\begin{eqnarray}
\Psi _{l,j=l-\frac{1}{2},m}\left( x\right)  &=&e^{-i\frac{Et}{\hbar }}r^{-%
\frac{1}{2}\left( 1-\frac{1}{b}\right) }  \nonumber \\
&&\times G_{\left( b\right) }\left( r\right) \left(
\begin{array}{c}
-\sqrt{\frac{l-m+\frac{1}{2}}{2l+1}}Y_{l}^{m-\frac{1}{2}}\left( \theta ,\phi
\right)  \\
\sqrt{\frac{l+m+\frac{1}{2}}{2l+1}}Y_{l}^{m+\frac{1}{2}}\left( \theta ,\phi
\right)
\end{array}
\right) ,  \label{300}
\end{eqnarray}
where $F_{\left( b\right) }\left( r\right) $ and $G_{\left( b\right) }\left(
r\right) $ are given by Eqs. (\ref{223}) and (\ref{222}), respectively.
It is important to call attention to the fact that all
 these results depends on the geometrical and
topological features of the global monopole spacetime.

\vskip 1.0 cm

\section*{IV. CONCLUSIONS}

With the purpose of
discussing the role of the topology on an  atomic system
we carried out the calculations of the shifts in the energy levels
of hydrogen atom placed in the spacetimes of a string and a monopole,
adding, in this way, some new results to the interesting problem
 considered in seminal
papers by Parker and collaborators[4-8] about the  effects of gravitational
fields at the atomic level, but now from the geometrical and topological
points of view, instead of looking only for the local effects of the
curvature as in those earlier papers[4-8].

 The presence of a cosmic string changes the solution and shifts
the energy levels of a hydrogen atom as compared with the flat Minkowski
spacetime result. It is interesting to observe that these
shifts depend on the parameter that defines the angle deficit and
 vanish when the angle deficit vanishes.  These shifts
arise from the topological  features of the spacetime generated by
this defect.

In the case of the hydrogen atom in the spacetime of a global monopole,
the modifications in the solution and  the shifts in the energy levels
are due to the combined effects of the curvature and the nontrivial topology
determined by the  deficit solid angle associated with this
 spacetime. These shifts also vanish when the deficit solid angle vanishes.

 Both effects
can be thought of as a consequence of the topological influence of the
spacetime under consideration upon the hydrogen atom.

The decrease in the energy for the situations considered is only two orders of
magnitude less than the ratio between the fine structure splitting and the
energy of the ground state of the non-relativistic hydrogen atom and is of
the order of the Zeeman effect. Therefore, the modifications in the spectra
of the hydrogen atom due to the presence of the gravitational fields of a
string or a monopole are all measurable, in principle.

The obtained results show how the geometry and a nontrivial topology
influences the energy spectrum as compared with the flat spacetime case and
show how these quantities depend on the surroundings and their
characteristics. These results also show how the solutions are modified.

Therefore, the problem of finding how the energy spectrum of an atom placed
in a gravitational field is perturbed by this background has to take into
account not only the geometrical, but also the topological features of the
spacetimes under consideration. In other words, the behaviour
of an atomic system is determined not only by the curvature at the
position of the atom, but also by the topology of the background
spacetime.

\section*{Acknowledgments}

We acknowledge Conselho Nacional de Desenvolvimento Cient\'{\i}fico e
Tecnol\'{o}gico (CNPq) and Coordena\c{c}\~{a}o de Aperfei\c{c}oamento de
Pessoal de N\'{\i}vel Superior (CAPES-Program PROCAD) for partial financial
support.

\end{document}